\documentclass[pdflatex,sn-nature]{sn-jnl}

\usepackage{graphicx}%
\usepackage{multirow}%
\usepackage{amsmath,amssymb,amsfonts}%
\usepackage{amsthm}%
\usepackage{mathrsfs}%
\usepackage[title]{appendix}%
\usepackage{xcolor}%
\usepackage{textcomp}%
\usepackage{manyfoot}%
\usepackage{booktabs}%
\usepackage{algorithm}%
\usepackage{algorithmicx}%
\usepackage{algpseudocode}%
\usepackage{listings}%
\usepackage{natbib}%

\usepackage{newtxmath}
\usepackage{textcomp}
\usepackage{float}
\usepackage[normalem]{ulem}
\usepackage{verbatim}
\usepackage{lineno}
\usepackage{comment}
\usepackage{bm}
\usepackage{titletoc}
\usepackage{ragged2e}
\usepackage{siunitx}
\usepackage{gensymb}

\theoremstyle{thmstyleone}%
\theoremstyle{thmstyletwo}%

\theoremstyle{thmstylethree}%
\begin{document}
\raggedbottom
\title[Article Title]{Polarization Engineering of Second-Harmonic Generation in 3R-MoS$_2$ Waveguides}


\author[1]{\fnm{Renkang} \sur{Song}}
\equalcont{These authors contributed equally to this work.}

\author[1]{\fnm{Junbo} \sur{Xu}}
\equalcont{These authors contributed equally to this work.}

\author[1]{\fnm{Yanzhen} \sur{Yin}}
\equalcont{These authors contributed equally to this work.}

\author[1]{\fnm{Yu} \sur{Yin}}
\author[2]{\fnm{Xu} \sur{Jiang}}
\author[1]{\fnm{Zhichen} \sur{Zhao}}
\author[1]{\fnm{Lei} \sur{Zhou}}
\author[3]{\fnm{Jintian} \sur{Lin}}
\author[3]{\fnm{Gaozhong} \sur{Wang}}
\author[4]{\fnm{Vasily} \sur{Kravstov}}
\author[5]{\fnm{Kyoung-Duck} \sur{Park}}
\author[4]{\fnm{Ivan} \sur{Iorsh}}
\author[6]{\fnm{Yuerui} \sur{Lu}}
\author[7,8]{\fnm{Jun} \sur{Wang}}
\author[2]{\fnm{Guangwei} \sur{Hu}}
\author[1]{\fnm{Zhanshan} \sur{Wang}}

\author*[1]{\fnm{Di} \sur{Huang}}\email{idgnauh@tongji.edu.cn}

\author*[1]{\fnm{Tao} \sur{Jiang}}\email{tjiang@tongji.edu.cn}

\affil[1]{\orgdiv{MOE Key Laboratory of Advanced Micro-Structured Materials, Shanghai Frontiers Science Center of Digital Optics, Institute of Precision Optical Engineering, and School of Physics Science and Engineering}, \orgname{Tongji University}, \orgaddress{\city{Shanghai}, \country{China}}}

\affil[2]{\orgdiv{School of Electrical and Electronic Engineering}, \orgname{Nanyang Technological University}, \orgaddress{\city{Singapore}, \country{Singapore}}}

\affil[3]{\orgdiv{State Key Laboratory of Ultra-intense Laser Science and Technology}, \orgname{Shanghai Institute of Optics and Fine Mechanics, Chinese Academy of Sciences}, \orgaddress{\city{Shanghai}, \country{China}}}

\affil[4]{\orgdiv{School of Physics and Engineering}, \orgname{ITMO University}, \orgaddress{\city{Saint Petersburg}, \country{Russia}}}

\affil[5]{\orgdiv{Department of Physics, Department of Semiconductor Engineering}, \orgname{Pohang University of Science and Technology}, \orgaddress{\city{Pohang}, \country{Republic of Korea}}}

\affil[6]{\orgdiv{School of Engineering, ANU College of Systems and Society}, \orgname{The Australian National University}, \orgaddress{\city{Canberra}, \state{Australian Capital Territory}, \country{Australia}}}

\affil[7]{\orgdiv{Department of Interdisciplinary Photonics}, \orgname{Shanghai Institute of Optics and Fine Mechanics, Chinese Academy of Sciences}, \orgaddress{\city{Shanghai}, \country{China}}}

\affil[8]{\orgdiv{Center of Materials Science and Optoelectronics Engineering}, \orgname{University of Chinese Academy of Sciences}, \orgaddress{\city{Beijing}, \country{China}}}


\abstract{\justifying

Chip-scale nonlinear optics enables strong light-matter interactions within compact devices, serving as a fundamental platform for multifunctional integrated photonics from classical optical signal processing to quantum information technologies.
Transition metal dichalcogenide (TMDC) waveguides have recently emerged as a highly promising platform owing to their giant material nonlinearity and extended interaction lengths. 
To date, however, research has predominantly focused on conversion efficiency, leaving the mechanisms governing the polarization state of nonlinear signal largely unexplored. 
Here, we establish a comprehensive framework for engineering the polarization of second-harmonic generation (SHG) in 3R-MoS$_2$ waveguides. 
By synergizing polarization-resolved measurements with theoretical modeling, we reveal that the SHG polarization is determined by guided-mode interactions constrained by waveguide geometry and crystal symmetry, and further reshaped during propagation. 
We demonstrate that thickness-dependent guided-mode confinement and in-plane crystal symmetry provide robust, static control over SHG polarization, while propagation length offers a dynamic tuning knob for continuously tailoring the nonlinear output. 
Our findings provide a deterministic approach for on-chip polarization engineering, opening opportunities for reconfigurable nonlinear light sources and quantum photonic circuits.

}

\maketitle
\newpage
\section{Introduction}\label{sec1}
Nonlinear optical processes are fundamental to cutting-edge photonics and quantum technologies \cite{shen2003principles, Boyd2008NonlinearOptics}, enabling critical functionalities such as frequency conversion \cite{chen2017enhanced, Xie2020On-Chip, rehan2025second, wu2023generation}, ultrafast optical switching \cite{feng2021all, Luo2024Strong, Sortino2025Atomic-layer, seidt2025ultrafast}, and quantum light generation \cite{Ma2020Ultrabright,  guo2023ultrathin, Fan2025Enhanced, feng2024polarization}.
Recently, two-dimensional (2D) van der Waals materials have opened new opportunities for nonlinear waveguides by combining large optical nonlinearities with atomically flat interfaces, customizable heterostructures, and broad electrical and structural tunability \cite{Zhang2009Direct, LAM2023060038, novoselov20162d, mak2016photonics, jiang2014valley, xia2014two, Ju2016Topological, Jiang2018Gate-tunable, hu2019coherent, Hong2023Twist-phase-matching}.
Among these, multilayer transition metal dichalcogenides (TMDCs) such as 3R-MoS$_2$ have emerged as a promising platform for nonlinear optical waveguides, combining long interaction length with strong nonlinearity for enhanced guided-wave frequency conversion \cite{xu2022towards, xu2025spatiotemporal, trovatello2025quasi}.

While previous studies on TMDC-based nonlinear photonics have primarily focused on improving conversion efficiency \cite{Li2025Efficient, Lin2026Nonlinear, tang2024quasi, tognazzi2025interface, peng20253r, abtahi2025thickness}, much less attention has been paid to the polarization characteristics of the nonlinear optical responses.
Beyond conversion efficiency, polarization represents a fundamental degree of freedom in photonic systems, forming the basis of essential functionalities such as information encoding \cite{Li2012Three-dimensional,Guo2021Full‐Color,Lu2025Atomic-scale,jin2025vectorial}, multiplexed communication \cite{Chen2016Use,Xiong2023Breaking,Wang2024Unlocking,Feng2025Bilayer}, and quantum-state manipulation \cite{Yin2017Satellite-Based,oes-2024-0024}.
However, in TMDC-based nonlinear waveguides, exemplified by 3R-MoS$_2$, the second-harmonic generation (SHG) polarization response is governed by the intricate coupling among waveguide modes constrained by the crystal symmetry, making its evolution complex to predict or control.

In this work, we establish a framework for understanding and engineering the polarization state of guided-wave SHG in 3R-MoS$_2$ waveguides.
By combining polarization-resolved measurements with symmetry-based modeling, we show that the generated polarization state results from the coupled effects of mode confinement, crystal symmetry, and propagation length along the waveguide.
These insights enable comprehensive manipulation of polarized nonlinear emission in integrated photonics, positioning 3R-MoS$_2$ waveguides as a promising platform for future optoelectronic and quantum information systems.

\section{Results}\label{sec2}

\begin{figure}[!htbp]
    \centering
    \includegraphics[width=1\textwidth]{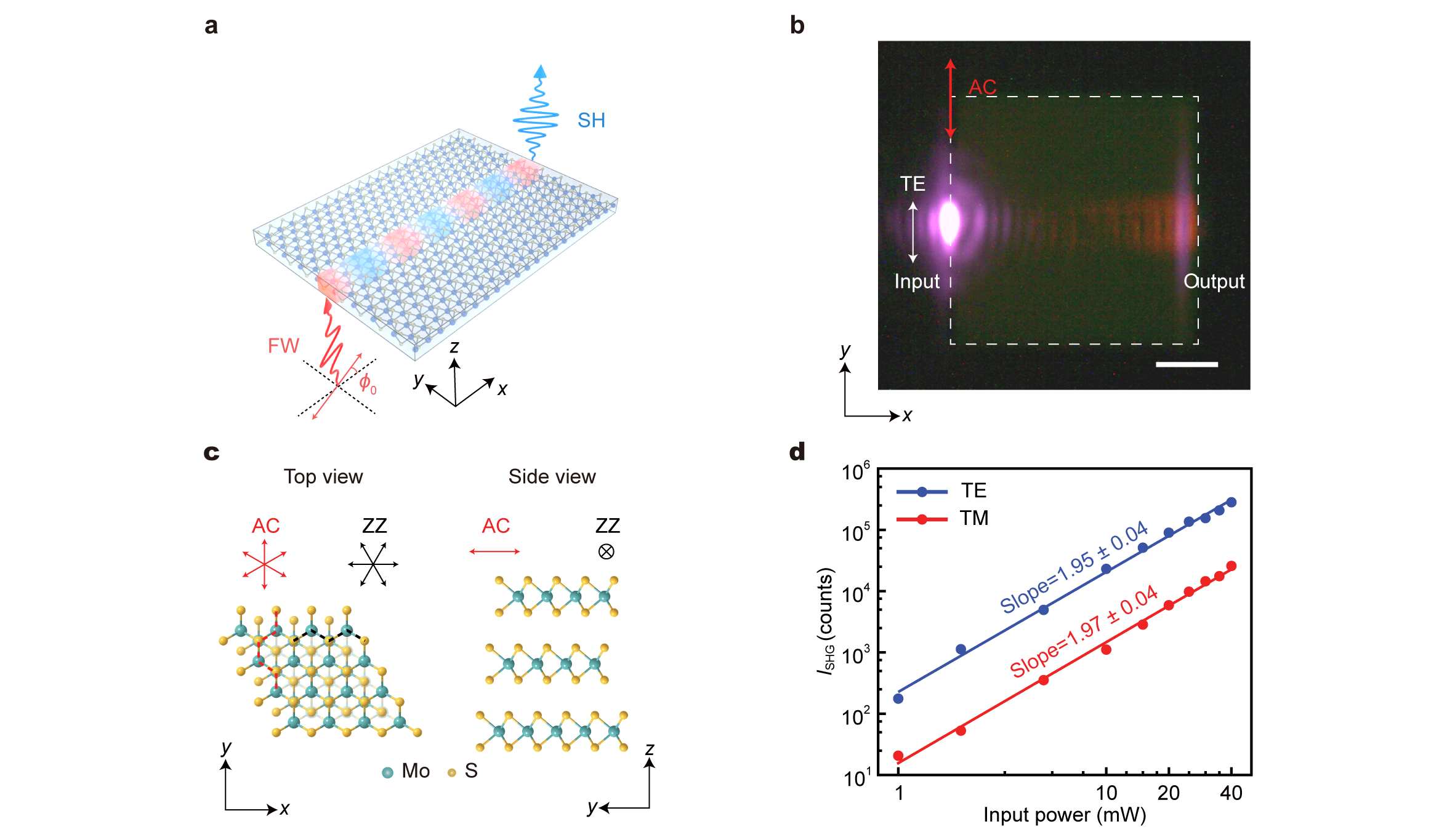}
\caption{\textbf{SHG in 3R-MoS$_2$ nonlinear TMDC waveguides.}
\textbf{a} Schematic of the edge-coupling measurement.
A linearly polarized femtosecond FW beam is focused onto the input edge of a 3R-MoS$_2$ waveguide to excite waveguide modes via edge scattering.
The incident polarization angle $\phi_0$ is defined with respect to the propagation direction ($x$-axis).
The generated SH waveguide modes co-propagate with the FW and are out-coupled at the output edge for polarization-resolved detection.
The coordinate system is defined with the $x$-axis along the light propagation direction, the $y$-axis parallel to the input edge, and the $z$-axis perpendicular to the substrate.
\textbf{b} Microscopic image of a fabricated 3R-MoS$_2$ waveguide ($\SI{357}{\nm}$ thick) on a fused silica (SiO$_2$) substrate, overlaid with spatially resolved false-colour images of the FW ($\lambda = \SI{1200}{\nm}$, magenta) and SH ($\lambda = \SI{600}{\nm}$, orange) signals under TE excitation.
Dashed lines outline the waveguide boundaries.
Scale bar: $\SI{5}{\um}$.
\textbf{c} Crystal structure of 3R-MoS$_2$ corresponding to the waveguide orientation in (\textbf{b}).
Left: top view in the $x$--$y$ plane, with red and black arrows (and dashed lines) indicating the AC and ZZ crystallographic directions, respectively.
Right: side view illustrating the 3R stacking sequence.
\textbf{d} SHG intensity ($I_{\mathrm{SHG}}$) measured at the output edge of the waveguide in (\textbf{b}) as a function of excitation power under TE and TM excitation, respectively 
}
\label{fig:fig1}
\end{figure}

To investigate the polarization state of SHG in 3R-MoS$_2$ waveguides, we employ an edge-coupling configuration in which waveguide modes are excited at the input edge and the generated SHG is detected at the output edge (Fig.~\hyperref[fig:fig1]{1a}).
A linearly polarized femtosecond fundamental-wave (FW) beam is focused onto the input edge, where edge coupling launches the FW beam into waveguide modes \cite{xu2022towards}.
During propagation, the intrinsic second-order nonlinearity of 3R-MoS$_2$ generates co-propagating second-harmonic (SH) waveguide modes.
At the output edge, both the FW and SH waveguide modes are out-coupled into free space and collected by a high-numerical-aperture objective lens \cite{zhu2017ultra} for polarization-resolved analysis.

The exfoliated 3R-MoS$_2$ flakes are patterned into square waveguides with edges intentionally aligned along selected crystallographic directions, where the in-plane crystallographic direction is determined beforehand by \textit{in-situ} co-polarized SHG measurements (Supplementary Information Sec.~1).
A representative $\SI{20}{\um}\times\SI{20}{\um}$ waveguide is shown in Fig.~\hyperref[fig:fig1]{1b}, with the input edge aligned parallel to the armchair (AC) axis, hereafter referred to as AC alignment.
For comparison, waveguides with the input edge aligned along the zigzag (ZZ) axis are referred to as ZZ alignment in the following.
Atomic force microscopy (AFM) measurements (Fig.~S1) confirm a uniform thickness of $\SI{357}{\nm}$ and an atomically flat surface for this waveguide, indicating negligible surface-roughness-induced scattering and enabling low-loss optical propagation over extended distances \cite{lee2023wafer}.
The corresponding lattice structure of 3R-MoS$_2$ including the crystallographic direction and the stacking sequence are illustrated in Fig.~\hyperref[fig:fig1]{1c}.

We first characterize the basic nonlinear optical response of the waveguide.
The incident FW polarization is set parallel to the input edge, corresponding to polarization angles \(\phi_0 = 90^\circ\) and \(270^\circ\), to predominantly excite transverse-electric (TE) waveguide modes. 
In contrast, a polarization perpendicular to the input edge, corresponding to \(\phi_0 = 0^\circ\) and \(180^\circ\), primarily excites transverse-magnetic (TM) modes.
Power-dependent measurements of the signal collected at the output edge exhibit near-quadratic scaling with the excitation power, with fitted exponents of $1.95 \pm 0.04$ for TE excitation and $1.97 \pm 0.04$ for TM excitation (Fig.~\hyperref[fig:fig1]{1d}), confirming the SHG nature of the observed signal.
We also observe pronounced interference fringes in the real-space images (Fig.~\hyperref[fig:fig1]{1b}), which are attributed to interference between forward-propagating and back-reflected FW waveguide modes, with the generated SH signals leaking into the far-field \cite{shi2022steering}.

\begin{figure}[!htbp]
    \centering
    \includegraphics[width=1\textwidth]{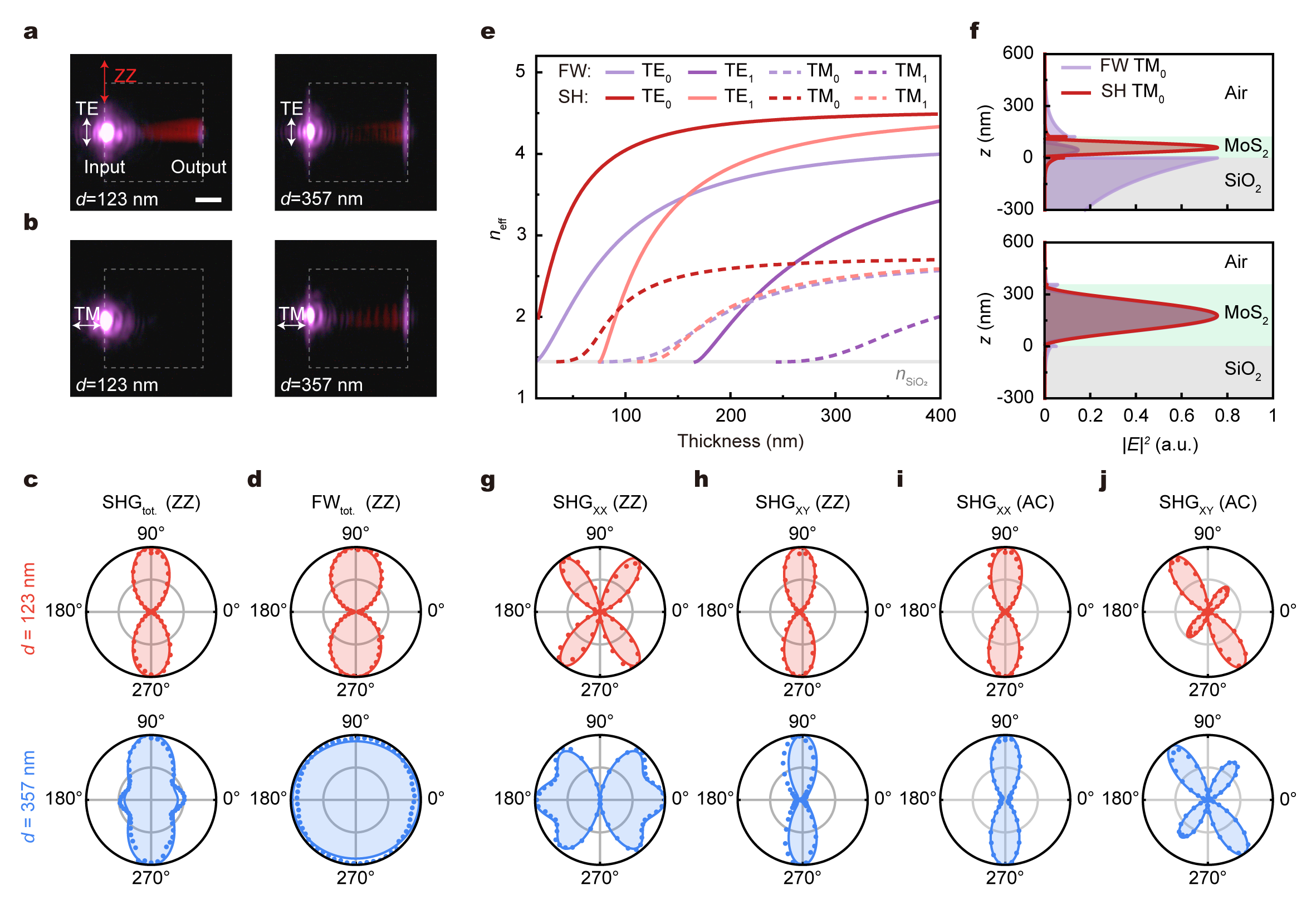}
    \caption{\textbf{Thickness-dependent polarization characteristics in 3R-MoS$_2$ waveguides.}
    \textbf{a}, \textbf{b} Real-space images of the FW (magenta) and SH (orange) signals in $\SI{20}{\um}\times\SI{20}{\um}$ square waveguides with thicknesses of $\SI{123}{\nm}$ and $\SI{357}{\nm}$ under TE (\textbf{a}) and TM (\textbf{b}) excitation, respectively. The input edges are aligned along the ZZ crystallographic direction. Scale bar: $\SI{5}{\um}$.
    \textbf{c}, \textbf{d} Polar plots of the normalized total SH (\textbf{c}) and FW (\textbf{d}) emission intensity as a function of the incident polarization angle $\phi_0$ for $\SI{123}{\nm}$- (red) and $\SI{357}{\nm}$-thick (blue) waveguides.
    \textbf{e} Thickness dependence of the effective refractive indices of the TE and TM modes at $\lambda=\SI{1200}{\nm}$ (FW) and $\lambda=\SI{600}{\nm}$ (SH), respectively.
    The horizontal gray line indicates the refractive index of the SiO$_2$ substrate ($n_{SiO_2} = 1.45$).
    \textbf{f} Simulated mode profiles of the fundamental TM$_0$ mode for FW and SH in $\SI{123}{\nm}$- and $\SI{357}{\nm}$-thick waveguides.
    \textbf{g}--\textbf{j} Polarization-resolved SHG emission under XX and XY detection for waveguides of different thicknesses with the input edge aligned along the ZZ axis (\textbf{g}, \textbf{h}) and the AC axis (\textbf{i}, \textbf{j}), respectively.
    In (\textbf{c}), (\textbf{d}), and (\textbf{g}--\textbf{j}), dots denote experimental data and solid lines are fits
    }
   \label{fig:fig2}
\end{figure}

The polarization state of the emitted SH field is dictated by the interplay of TE and TM modes supported at the FW and SH wavelengths. 
Since the confinement and availability of these modes are highly sensitive to waveguide geometry, waveguide thickness serves as a primary degree of freedom for manipulating the polarization of nonlinear signal. 
We therefore initiate our study by probing the thickness-dependent SHG evolution, starting with waveguides under ZZ alignment.
Figures~\hyperref[fig:fig2]{2a, b} compare the real-space FW an
d SHG signals for two representative thicknesses, $\SI{123}{nm}$ and $\SI{357}{nm}$.
Under TE excitation, both the transmitted FW and the generated SHG are clearly observed at the output edge for both thicknesses.
In contrast, under TM excitation, the SHG remains pronounced in the 357-nm-thick waveguide but is strongly suppressed in the 123-nm-thick waveguide, revealing a pronounced thickness dependence of the TM-excited response.

To quantitatively characterize this thickness-dependent contrast observed in the real-space images, we measure the SH intensity as a function of the incident polarization angle $\phi_0$.
As shown in Fig.~\hyperref[fig:fig2]{2c}, the 123-nm-thick waveguide exhibits a dominant two-lobe SHG pattern maximized under TE excitation ($\phi_0 = 90^\circ$ and $270^\circ$), with negligible response near $\phi_0 = 0^\circ$ and $180^\circ$ corresponding to TM excitation.
By contrast, the thicker waveguide ($\SI{357}{nm}$) develops additional lobes at $\phi_0=0^\circ$ and $180^\circ$, producing a clear four-lobe pattern that indicates a strong SHG response under TM excitation.

The FW response (Fig.~\hyperref[fig:fig2]{2d}) provides direct insight into the physical origin of this thickness dependence.
For the 123-nm-thick waveguide, the FW output is strongly anisotropic, with a significantly suppressed TM-excited FW signal.
This behavior is consistent with the calculated modal dispersion (Fig.~\hyperref[fig:fig2]{2e}), in which the FW TM$_0$ mode lies close to cutoff with an effective index approaching the refractive index of the SiO$_2$ substrate.
As a result, the reduced index contrast leads to weak field localization within the 3R-MoS$_2$ layer, diminished edge coupling, and enhanced radiative leakage during propagation.
In addition, the TM modes at the FW and SH wavelengths exhibit mismatched spatial profiles (Fig.~\hyperref[fig:fig2]{2f}, top panel), which reduces the nonlinear modal overlap \cite{reverse_stack}.
Together, these effects suppress the TM-excited SHG in the 123-nm-thick waveguide.

In contrast, for the $\SI{357}{nm}$-thick waveguide, the FW polarization response becomes much less anisotropic, with a substantially stronger TM-excited FW output (Fig.~\hyperref[fig:fig2]{2d}, bottom panel), consistent with improved confinement of the TM mode.
More efficient TM-mode excitation, reduced radiative loss, and enhanced nonlinear modal overlap enabled by better-confined fields at both FW and SH wavelengths (Fig.~\hyperref[fig:fig2]{2f}, bottom panel) collectively lead to a pronounced TM-excited SHG response.
Waveguides with intermediate thicknesses ($\SI{225}{nm}$ and $\SI{332}{nm}$) exhibit the same trend (Supplementary Information Sec.~4), supporting a progressive strengthening of TM-excited SHG as modal confinement and nonlinear modal overlap improve with increasing thickness.

While the above measurements establish a strong thickness dependence of the SHG intensity---reflecting the role of modal confinement and nonlinear modal overlap---they do not yet reveal how the polarization of the generated SH fields evolves with thickness. 
We therefore further analyze the thickness-dependent polarization of SH signal under co-polarized (XX) and cross-polarized (XY) configurations with respect to the incident FW polarization \cite{li2013probing}. 
The polarization-resolved SHG results reveal a clear thickness dependence in the XX and XY patterns (Fig.~\hyperref[fig:fig2]{2g, h}).
For the thicker 357-nm-thick waveguide, the enhancement of the TM-excited SHG manifests predominantly in the XX pattern, while it does not appear in the XY pattern.
This comparison indicates that the thickness-dependent increase in TM-excited SHG arises primarily from a TM component.
By contrast, in the 123-nm-thick waveguide, the TM-excited SHG remains weak in both XX and XY patterns, consistent with the inefficient TM-mode excitation and reduced nonlinear modal overlap discussed above.

To examine whether this thickness-dependent polarization behavior is influenced by crystal orientation, we adjust the input-edge alignment to the AC axis (Fig.~\hyperref[fig:fig2]{2i, j}).
In this configuration, the TM-excited XX signal is strongly suppressed, even in the 357-nm-thick waveguide.
Such suppression originates from crystal-symmetry constraints that forbid the in-plane nonlinear interaction responsible for generating the TM-polarized SH component under AC alignment (Supplementary Information Sec.~5).
This observation indicates that even when TM-polarized SHG is enabled by sufficient waveguide thickness, crystal orientation can selectively suppress specific polarization components through symmetry constraints.

More generally, the polarization-resolved SHG response exhibits a pronounced dependence on crystal orientation that persists for both thin and thick waveguides.
Specifically, the SHG polarization patterns of waveguides under ZZ alignment remain largely unchanged when the incident linear polarization angle is reversed from $\phi_0$ to $-\phi_0$, whereas waveguides under AC alignment exhibit pronounced asymmetry, most clearly observed in the XY pattern (Fig.~\hyperref[fig:fig2]{2j}).
This contrast originates from how crystal symmetry governs the interference between nonlinear contributions with different responses to the polarization-angle reversal, thereby determining whether these contributions remain effectively separated or interfere in the detected SHG polarization component (for more details, see Supplementary Information Sec.~5).
These results show that while waveguide thickness critically governs the strength of TM-polarized SHG through modal confinement, the in-plane crystal orientation plays an equally essential role in shaping the polarization of nonlinear response.

\begin{figure}[!htbp]
    \centering
    \includegraphics[width=1\textwidth]{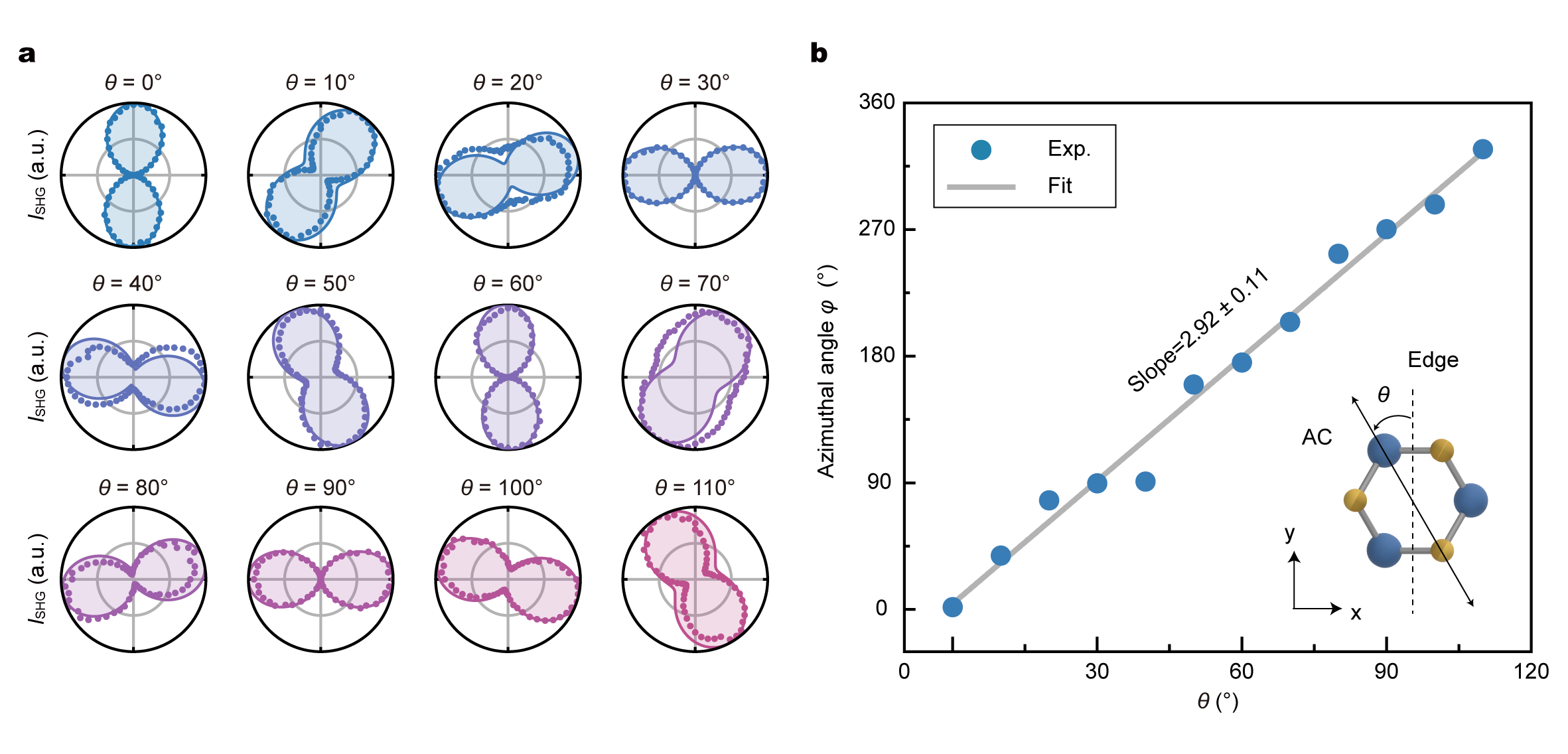}
    \caption{\textbf{Crystal-orientation control of SHG polarization.}
    \textbf{a} Polar plots of the normalized SHG polarization emission for $\theta$ ranging from $0^\circ$ to $110^\circ$ in $10^\circ$ steps under TE excitation.
    Dots: experimental data; solid curves: fits using Eq.~\ref{eq:shg_field}.
    \textbf{b} Extracted SHG polarization azimuthal  angle $\varphi$ as a function of the crystal orientation angle $\theta$ (inset: definition of crystal orientation angle $\theta$).
    Symbols: experimental data; solid line: linear fit, yielding a slope of $2.92 \pm 0.11$ and indicating a near-$3\theta$ dependence of $\varphi$
   }
    \label{fig:fig3}
\end{figure}

We then systematically investigate this orientation dependence using square 3R-MoS$_2$ waveguides of uniform thickness ($\SI{225}{\nm}$) and identical lateral dimensions ($\SI{13.5}{\um} \times \SI{13.5}{\um}$), fabricated with different in-plane crystal orientations.
The orientation is parameterized by the angle $\theta$ between the AC axis and the input edge of the waveguide.
Owing to the threefold rotational symmetry of 3R-MoS$_2$, it is sufficient to characterize $\theta$ over a $120^\circ$ range.
Accordingly, we investigate 12 orientations from $\theta=0^\circ$ to $110^\circ$ in $10^\circ$ steps (see Supplementary Fig.~S12 for experimental details).
Under TE excitation, the output SHG polarization exhibits a pronounced dependence on $\theta$ in Fig.~\hyperref[fig:fig3]{3a}.
For each orientation, we extract the SH polarization azimuthal $\varphi$ by fitting the analyzer-angle dependence with a standard sinusoidal form for an elliptically polarized field.
The resulting $\varphi(\theta)$ exhibits an overall near-$3\theta$ dependence (Fig.~\hyperref[fig:fig3]{3b}).

To capture the crystal-orientation dependence in a unified manner, we develop a symmetry-based model that explicitly links the edge-emitted SH field to the crystal orientation $\theta$ (see Supplementary Information Sec.~6).
Accordingly, the in-plane components of the edge-emitted SH field under TE excitation can be expressed as:
\begin{equation}
\begin{pmatrix}E_x^{2\omega} \\ E_y^{2\omega}\end{pmatrix}=
\begin{pmatrix}-\alpha \sin(3\theta) + \gamma \\ \beta \cos(3\theta)\end{pmatrix}
\label{eq:shg_field}
\end{equation}
here $\alpha$ and $\beta$ represent the in-plane nonlinear contributions and $\gamma$ captures the out-of-plane contribution.
Unlike the sinusoidal fits used above to retrieve $\varphi$ for each individual orientation, Eq.~\ref{eq:shg_field} imposes the $C_{3v}$-symmetry constraint on the $\theta$ dependence of the SH field and therefore describes all measured polarization patterns with a single parameter set.
Using one set of effective coefficients $(\alpha,\beta,\gamma)$, Eq.~\ref{eq:shg_field} reproduces the polarization-resolved patterns over the full range of $\theta$ in Fig.~\hyperref[fig:fig3]{3a}, validating the symmetry-based description and the crystal-orientation correlation captured by the model.
In the ideal limit where $\gamma \rightarrow 0$ and $\alpha \simeq \beta$, Eq.~\ref{eq:shg_field} yields $\varphi=3\theta$, consistent with the experimentally observed near-$3\theta$ dependence.

Importantly, while the crystal orientation $\theta$ sets the symmetry-imposed angular dependence of the SHG polarization, as captured by Eq.~\ref{eq:shg_field}, the experimentally observed deviations from an ideal $3\theta$ dependence (Fig.~\hyperref[fig:fig3]{3b}) demonstrate that the polarization state is also critically influenced by the relative amplitudes and phases of the TE- and TM-polarized SH components, encoded in the coefficients $\alpha$, $\beta$, and $\gamma$.
Unlike the crystal orientation, these coefficients depend not only on purely geometric parameters, but also on modal properties and phase relations, suggesting the polarization evolution during propagation.
Indeed, measurements at other pump wavelengths (e.g., $\lambda = \SI{1500}{nm}$) reveal systematic deviations from an ideal $3\theta$ dependence, reflecting the influence of modal and propagation effects, while remaining fully consistent with the underlying symmetry constraints (Supplementary Information Sec.~6).

\begin{figure}[!htbp]
    \centering
    \includegraphics[width=1\textwidth]{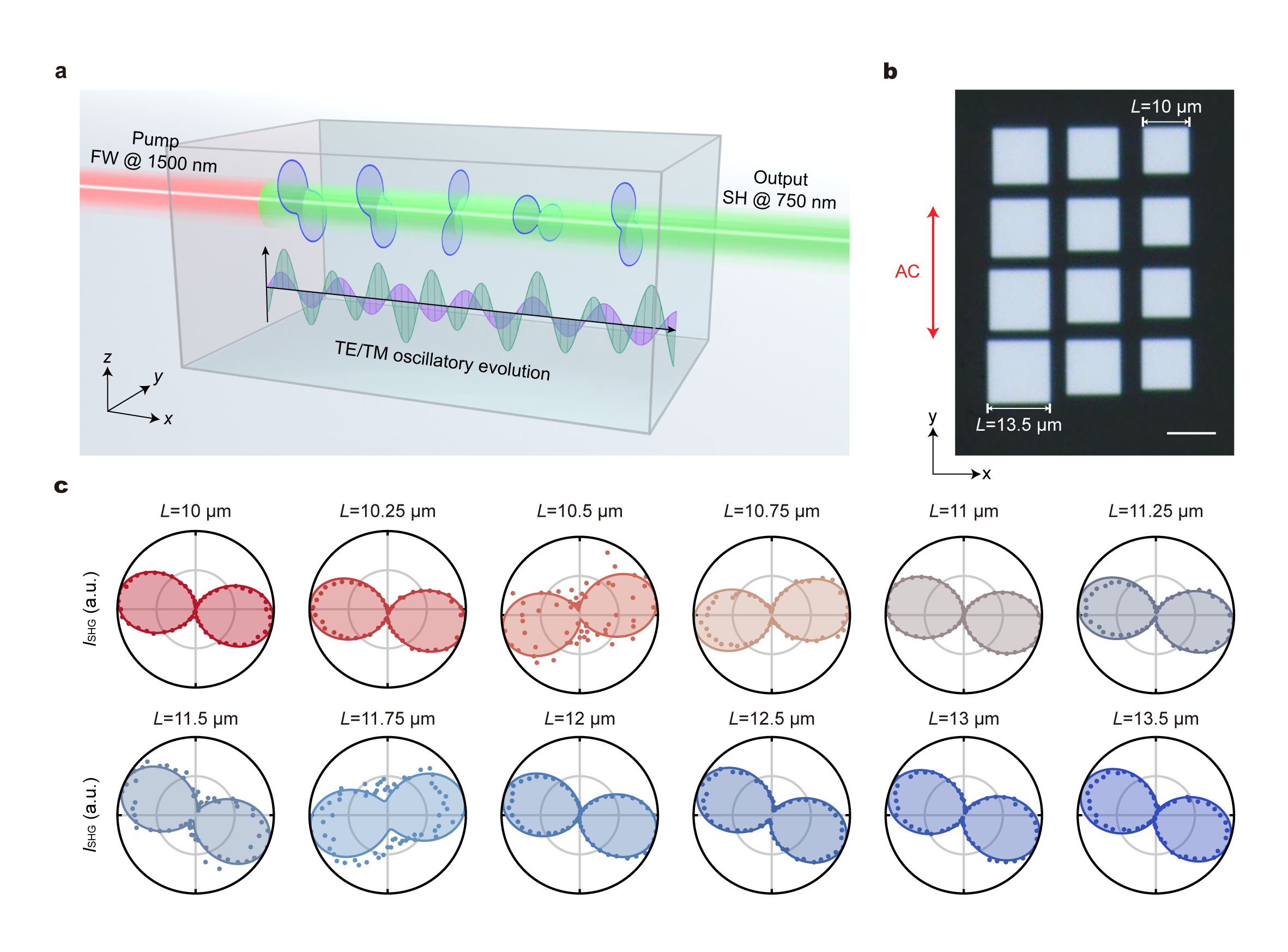}
    \caption{\textbf{Propagation-induced evolution of SHG polarization.}
    \textbf{a} Schematic illustration of SHG polarization evolution. 
    FW waveguide modes generate co-propagating SH fields, and the edge-emitted output polarization evolves continuously with propagation length due to interference among SH fields generated along the waveguide and the accumulated relative phase delay between SH TE and TM components.
    \textbf{b} Optical micrograph of a set of $\SI{225}{\nm}$-thick 3R-MoS$_2$ waveguides ($\theta=0^\circ$) with systematically varied lengths $L$. Scale bar: $\SI{10}{\um}$.
    \textbf{c} Polar plots of the edge-emitted SHG polarization for $L$ from $\SI{10}{\um}$ to $\SI{13.5}{\um}$ under $\phi_0=45^\circ$ excitation (dots: data; solid lines: fits)
    }
    \label{fig:fig4}
\end{figure}

While the above results establish waveguide thickness and crystal orientation as static control knobs of SH polarization, they do not capture the dynamic evolution of the generated SH fields during propagation.
In the 3R-MoS$_2$ waveguide, the generated SH fields undergo propagation-induced evolution, with both their relative amplitudes and phases varying along the propagation direction.
This naturally motivates an investigation of how the SHG polarization evolves with propagation length $L$ (Fig.~\hyperref[fig:fig4]{4a}).


To investigate the effect of propagation length, we fabricate $\SI{225}{\nm}$-thick 3R-MoS$_2$ waveguides with different lengths (Fig.~\hyperref[fig:fig4]{4b}).
The pump wavelength is set to \SI{1500}{nm} to reduce excitonic absorption at the SH wavelength, the incident polarization is chosen as $\phi_0 = 45^\circ$ to excite both TE and TM FW modes, and the crystal orientation is fixed at $\theta = 0^\circ$.
Figure~\hyperref[fig:fig4]{4c} reveals a pronounced evolution of the edge-emitted SHG polarization with increasing propagation length $L$. As $L$ is varied from \SI{10}{\um} to \SI{13.5}{\um}, the output polarization state continuously changes in both azimuthal  angle $\varphi$ and ellipticity, demonstrating that the SHG polarization is not fixed at generation but evolves dynamically during propagation.
Notably, these changes occur under otherwise identical conditions, indicating that the observed polarization evolution is governed by propagation-induced amplitude oscillation and phase accumulation between multiple TE- and TM-polarized SH waveguide modes.

This behavior is consistent with the calculated evolution of the relative SH amplitudes and phase delay between the SH TE and TM components (Fig.~S13a, b).
When represented on the Poincar\'e sphere, the corresponding Stokes parameters follow a continuous trajectory as $L$ increases (Fig.~S13c), providing a visualization of the oscillatory polarization evolution induced by multimode interactions.
Importantly, the shape and orientation of this trajectory can be further engineered by adjusting waveguide geometry or crystal orientation, establishing propagation length as a dynamic control parameter for nonlinear polarization engineering in integrated photonic platforms.

\section{Discussion}\label{sec13}

In summary, we establish polarization as a critical and controllable degree of freedom in TMDC waveguides.
Moving beyond efficiency-centered metrics, our study shows that the polarization state of guided-wave SHG can be systematically controlled through the interplay between waveguide geometry, crystal symmetry, and propagation-induced evolution of the relative amplitudes and phase between SH waveguide modes.
This hierarchy of control parameters allows the platform to function as a ``nonlinear on-chip waveplate", enabling continuous and deterministic shaping of the polarization of nonlinear output. By transforming the waveguide from a simple frequency converter into a nonlinear polarization shaper, our work provides a deterministic approach for on-chip polarization engineering. 
More broadly, the concepts demonstrated here are readily extendable to other van der Waals materials and nonlinear processes, opening routes toward ultra-compact devices for polarization-division multiplexing, tunable quantum light sources, and integrated quantum photonic circuitry.

\backmatter

\section{Materials and Methods}
 
\noindent\textbf{Sample fabrication} 

\noindent The 3R-MoS$_2$ flakes are mechanically exfoliated from bulk crystals (HQ Graphene) onto a fused silica substrate that is first cleaned using a vacuum plasma cleaner (SUNJUNE PLASMA VPR3). 
The thickness of the exfoliated flakes is characterized by AFM. 
Microstructures with boundaries aligned along specific crystallographic orientations are defined by electron beam lithography (EBL) followed by reactive ion etching (RIE), resulting in square 3R-MoS$_2$ patterns. 
The patterned flakes are then transferred to an atomically flat fused silica substrate to minimize extra scattering from substrate roughness introduced during the etching process.\\

\noindent\textbf{Optical measurements} 

\noindent Waveguide emission is characterized in a confocal transmission setup (Fig.~S2). 
The femtosecond laser system (Levante IR fs, APE GmbH) is operated at center wavelengths of \SI{1200}{nm} and \SI{1500}{nm}.
The output is polarized using a Glan-Taylor polarizer and a half-wave plate, and is then focused onto the sample edge via a 100$\times$ objective (NA = 0.5).
The output is collected by a second 100$\times$ objective (NA = 0.7), passed through a $\SI{100}{\upmu m}$ pinhole for spatial filtering, and spectrally separated by bandpass filters before entering the spectrometer.
All measurements are conducted under ambient laboratory conditions.\\

\noindent\textbf{Numerical simulations} 

\noindent The waveguide is modeled as a planar 3R-MoS$_2$ slab on a SiO$_2$ substrate.
The modal characteristics, including the effective indices and field profiles of the TE and TM modes at the FW (\(\SI{1200}{nm}\)) and SH (\(\SI{600}{nm}\)) wavelengths, are derived analytically from Maxwell's equations (see Supplementary Information Sec.~2 for more details).
The complex refractive indices of 3R-MoS$_2$ and SiO$_2$ are extracted from the literature \cite{MoS2_permittivity,malitson1965interspecimen}.
The resulting nonlinear interaction between waveguide modes is then calculated by solving the nonlinear coupled-mode equations, as detailed in Supplementary Information Sec.~3.

\section{Data availability}
The rest of the data are available from the corresponding authors upon reasonable request.

\section{Acknowledgements}
The authors acknowledge support from the National Key R{\&}D Program of China (2024YFB2808100), the National Natural Science Foundation of China (62475194, 62305249, 62192770, 62192772, 12574364),  the Science and Technology Commission of Shanghai Municipality (23190712300, 23ZR1465800), and the Strategic Priority Research Program of the Chinese Academy of Sciences (Grant No. XDB0890000). \\ \\

\section{Author contributions}
R.S., J.X., and Y.Z.Y. contributed equally to this work. T.J. conceived and designed the experiments. Y.Z.Y., R.S., and Y.Y. fabricated the samples and conducted the far-field measurements. J.X. performed the calculations and simulations. R.S., Y.Z.Y., and J.X. contributed to the data analysis and manuscript writing with the help of all authors.
T.J. and D.H. supervised the entire project. All authors discussed and interpreted the results.

\section{Competing interests}
\noindent The authors declare no competing interests.

\section{Supplementary Information}
\textbf{This PDF file includes:}

Sections S1 to S7

Figs. S1 to S13

Table 1




\bigskip


\bibliography{Main-Ref}

\end{document}